\documentclass[runningheads]{llncs}

\usepackage[T1]{fontenc}
\usepackage[utf8]{inputenc}  
\usepackage{graphicx}
\usepackage{amsmath,amssymb}
\usepackage{booktabs}
\usepackage{multirow}
\usepackage{algorithm}
\usepackage{algpseudocode}
\usepackage{cite}
\usepackage{float}
\usepackage{makecell}
\usepackage[hidelinks]{hyperref}
\usepackage{marvosym} 

\begin{document}

\title{DSRAG: A Domain-Specific Retrieval Framework Based on Document-derived Multimodal Knowledge Graph}

\titlerunning{DSRAG: Domain-Specific Retrieval with Multimodal KG}

\author{
  Mengzheng Yang\inst{1} \and
  Yanfei Ren\inst{1}\textsuperscript{(\Letter)} \and
  David Osei Opoku\inst{2} \and
  Ruochang Li\inst{1} \and
  Peng Ren\inst{2} \and
  Chunxiao Xing\inst{2}
}

\authorrunning{M. Yang et al.}

\institute{
  School of Software, Henan University, Kaifeng 475004, China \\
  \email{renyanfei@henu.edu.cn} \and
  BNRist, DCST, RIIT, Tsinghua University, Beijing 100084, China
}


\maketitle

\begin{abstract}
Current general-purpose large language models (LLMs) commonly exhibit knowledge hallucination and insufficient domain-specific adaptability in domain-specific tasks, limiting their effectiveness in specialized question answering scenarios. Retrieval-augmented generation (RAG) effectively tackles these challenges by integrating external knowledge to enhance accuracy and relevance. However, traditional RAG still faces limitations in domain knowledge accuracy and context modeling.To enhance domain-specific question answering performance, this work focuses on a graph-based RAG framework, emphasizing the critical role of knowledge graph quality during the generation process. We propose DSRAG (Domain-Specific RAG), a multimodal knowledge graph-driven retrieval-augmented generation framework designed for domain-specific applications. Our approach leverages domain-specific documents as the primary knowledge source, integrating heterogeneous information such as text, images, and tables to construct a multimodal knowledge graph covering both conceptual and instance layers. Building on this foundation, we introduce semantic pruning and structured subgraph retrieval mechanisms, combining knowledge graph context and vector retrieval results to guide the language model towards producing more reliable responses. Evaluations using the Langfuse multidimensional scoring mechanism show that our method excels in domain-specific question answering, validating the efficacy of integrating multimodal knowledge graphs with retrieval-augmented generation.

\end{abstract}
\section{Introduction}\label{sec1}
\begin{figure}
        \centering
            \includegraphics[width=0.6\linewidth]{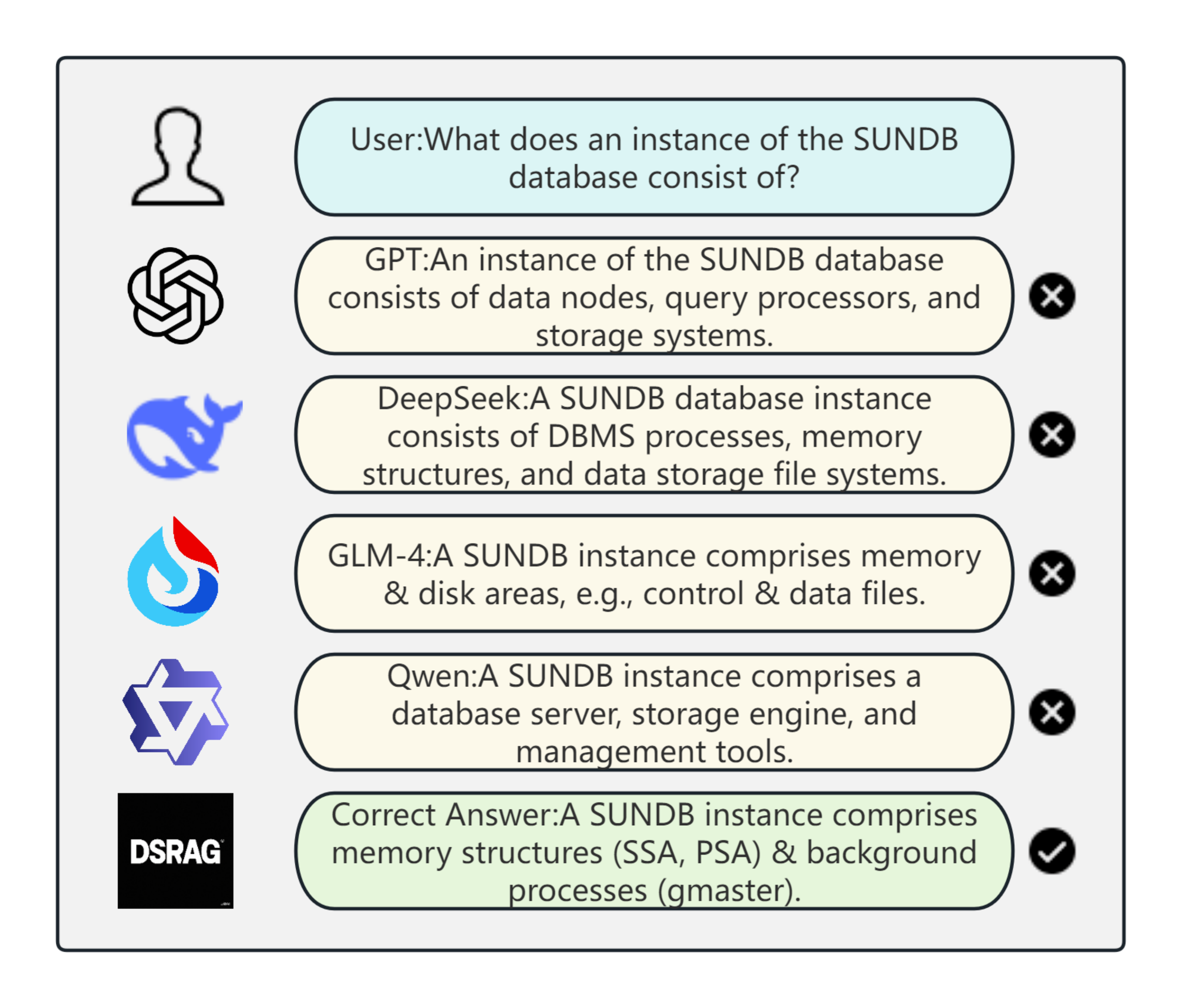}
            \caption{Diagram of the example of factual hallucination}
            \label{fig1}
        \end{figure}

	Current general-purpose large language models (LLMs) face two critical challenges in domain-specific tasks: firstly, there is a notable deficiency in the coverage of specialized knowledge in the pre-training data \cite{gu2021domain}; secondly, the outputs of the models often generate factual hallucinations and logical inconsistencies \cite{ji2023survey}. Models like ChatGPT, DeepSeek, GLM, and Qwen showcase excellent semantic understanding in general domains, but they exhibit notable limitations in domain-specific applications. As illustrated by the case in Fig.\ref{fig1}, when confronted with complex professional queries, LLMs frequently produce factual hallucinations \cite{huang2025survey}. This phenomenon exposes the cognitive gap between the data-driven statistical learning paradigm and structured knowledge, underscoring the urgent need to integrate external knowledge enhancement mechanisms into language models to enhance their reliability in domain-specific domains \cite{hu2023survey}.
    
	To address the knowledge gap in domain-specific domains, the academic community primarily explores two technical approaches: unsupervised incremental pre-training \cite{singhal2023large} and retrieval-augmented generation (RAG) \cite{lewis2020retrieval} mechanisms. The former enhances knowledge by injecting domain-specific corpora, but the high training costs and delayed knowledge update efficiency limit its applicability; the latter improves generation reliability through dynamic retrieval from external knowledge bases, yet still faces limitations in semantic understanding and precise retrieval \cite{barnett2024seven}.
	
	GraphRAG \cite{peng2024graph} has garnered widespread attention as an innovative solution. Unlike traditional RAG, GraphRAG introduces structured knowledge graphs to organize semantic units in a graph structure, thereby utilizing the semantic relationships between entities in the retrieval and generation processes. However, the performance of GraphRAG is highly dependent on the accuracy, completeness, and coverage of the knowledge graph. Especially in domain-specific domains, constructing high-quality knowledge graphs presents numerous challenges \cite{abu2021domain}. The primary carriers of domain-specific domain knowledge are multimodal documents, which contain information such as text, images, and tables. Therefore, constructing high-quality knowledge graphs from multimodal documents is a key prerequisite for achieving reliable Graph-based RAG.
	
	In response to these challenges, this paper proposes DSRAG, a multimodal knowledge graph-driven retrieval-augmented generation framework tailored for domain-specific tasks. We first leverage the document’s table of contents, along with large language models and expert annotations, to systematically generate summaries, extract domain concepts, and construct the concept knowledge graph. Then, based on the framework of the concept knowledge graph, we extract entities, relationships, and relevant data from the documents to construct the instance knowledge graph. Based on the multimodal knowledge graph, this paper further designs a graph-enhanced retrieval method, combining semantic subgraph retrieval and vector matching, thereby significantly improving the accuracy and efficiency of domain-specific question-answering tasks. 

\bigskip
\noindent\textbf{Contributions.} The main contributions are summarized as follows:

\begin{enumerate}
    \item We propose DSRAG, a framework designed for domain-specific tasks, combining multimodal knowledge graphs with retrieval-augmented generation.

    \item We construct the DSKG multimodal knowledge graph, which integrates text, images, and tables, encompassing both conceptual and instance layers for systematic modeling and high-quality knowledge foundation in domain-specific question answering.

    \item We introduce DSKG-Enhanced Retrieval, a graph-based retrieval method that combines structured semantic retrieval with vector matching, enhancing the accuracy and robustness of RAG in domain-specific question answering.
\end{enumerate}

    	\section{Related Work}\label{Work2}
	Traditional retrieval-augmented generation (RAG) methods primarily rely on external knowledge bases to enhance the performance of generation models. The RAG method proposed by Lewis et al. achieved significant results in open-domain tasks \cite{lewis2020retrieval}. However, traditional RAG methods exhibit certain limitations when handling structured semantic representations. To overcome this issue, Graph-based RAG (GraphRAG) has emerged. GraphRAG \cite{peng2024graph} introduces knowledge graphs to organize semantic units in a graph structure, enabling the retrieval phase to leverage relationships between entities to improve relevance matching, and enhancing the semantic constraints of the model during generation by incorporating graph information, thereby reducing hallucinations and logical conflicts. Compared with traditional RAG methods, GraphRAG places greater emphasis on leveraging entities and relationships within structured knowledge graphs to improve the accuracy and interpretability of generation models.
    
	Knowledge graph is the core component of GraphRAG, organizing and storing domain knowledge through a graph structure. Traditional knowledge graph construction methods mostly rely on structured data, focusing primarily on entity and relation extraction \cite{suchanek2007yago}, but they face limitations when handling multimodal data, especially in technical documents where images, tables, and other non-textual information are equally important \cite{xu2020layoutlm}. With the growing demand for multimodal data, an increasing number of studies have begun to explore how to integrate different types of data to enhance the construction of knowledge graphs \cite{zhu2022multi}. For example, the Mukea method \cite{ding2022mukea} proposed by Ding et al . enhances the performance of visual question answering systems by integrating image and text data, while the MORLD platform \cite{sheng2025mqrld} proposed by Sheng et al. improves the efficiency of multimodal data retrieval. However, these methods still face challenges in efficiently integrating data sources within domain-specific scenarios.
	
	The proposal and application of GraphRAG have gradually become a research hotspot. HybridRAG \cite{sarmah2024hybridrag} enhances the handling of domain-specific terminology and complex document formats by combining knowledge graphs with vector retrieval. In parallel, DO-RAG \cite{opoku2025dorag} focuses on domain-specific question answering by integrating knowledge graph-enhanced retrieval to improve the quality and relevance of generated answers.In addition, companies such as PingCAP and InfiniFlow have conducted related research, proposing AutoFlow \cite{pingcap_tidb_autoflow_2023} and RAGFlow \cite{infiniflow_ragflow_2024}, respectively. However, current graph-based RAG methods still face several challenges when applied to multimodal domain-specific documents: (1) the lack of systematic modeling of the conceptual and instance layers makes it difficult to form a structured knowledge system from abstraction to specificity; (2) the failure to fully integrate multiple information sources such as text, images, and tables limits the expressiveness of the knowledge graph; (3) due to the incomplete construction of the knowledge graph, a large amount of noisy information is introduced during retrieval, affecting the quality of generation.
	
	To address these challenges, this paper proposes an innovative method called DSRAG, which enhances RAG's retrieval and generation capabilities by constructing a high-quality knowledge graph covering both the conceptual and instance layers from domain-specific multimodal documents. This method integrates multimodal information such as text, images, and tables to systematically model the structure of domain knowledge, aiming to improve the reliability of domain-specific question answering.

	\section{Methodology}\label{Work3}
    \subsection{Framework of DSRAG}
\begin{figure}
    \centering
    \includegraphics[width=0.6\linewidth]{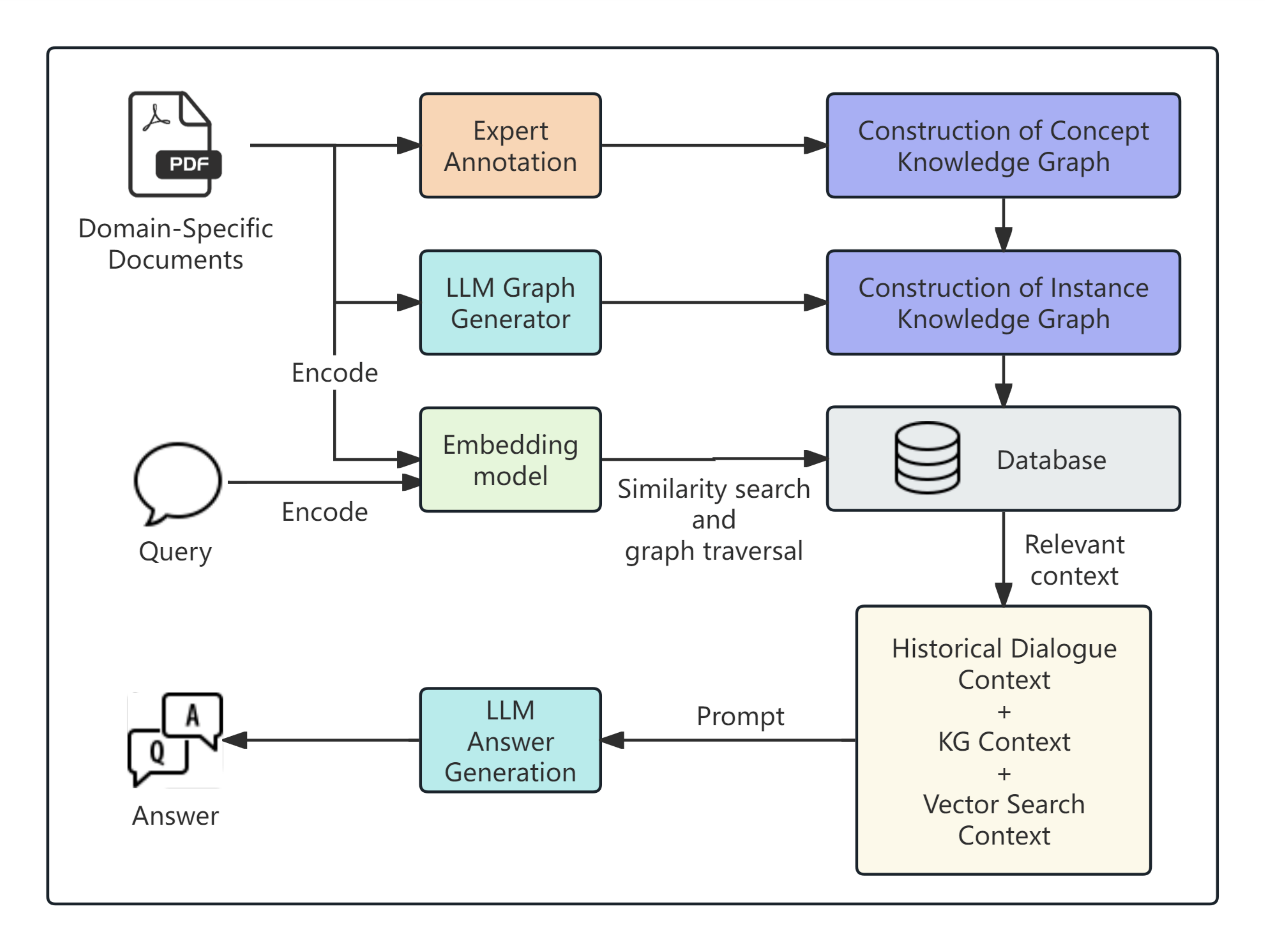} 
    \caption{The overall framework of DSRAG}
    \label{fig:second-image} 
\end{figure}

	To address semantic heterogeneity and factual hallucinations in domain-specific question answering, this paper introduces DSRAG, a retrieval-augmented generation framework driven by a domain-specific multimodal knowledge graph (DSKG), as shown in Fig. \ref{fig:second-image}. Unlike traditional RAG methods, DSRAG constructs a hierarchical knowledge graph by processing structured domain documents to build both the Concept KG and Instance KG.The Concept KG models core domain concepts and workflows by extracting summaries and keywords from the document's structure, combining large language models (LLMs) with expert annotations. The Instance KG constructs fine-grained triple representations by extracting entities and relations from both text and multimodal data.
    
    The DSKG-Enhanced Retrieval process further improves accuracy through a two-stage retrieval strategy: the first stage retrieves structured semantic information via graph traversal, while the second stage retrieves contextually relevant information through vector similarity search. Finally, the graph-based context, vector retrieval context, and historical dialogue context are integrated into a prompt, which is fed into the LLM to generate the final answer, ensuring both factual accuracy and semantic consistency.

    \subsection{DSKG Construction}
\begin{figure}
    \centering
    \includegraphics[width=1\linewidth]{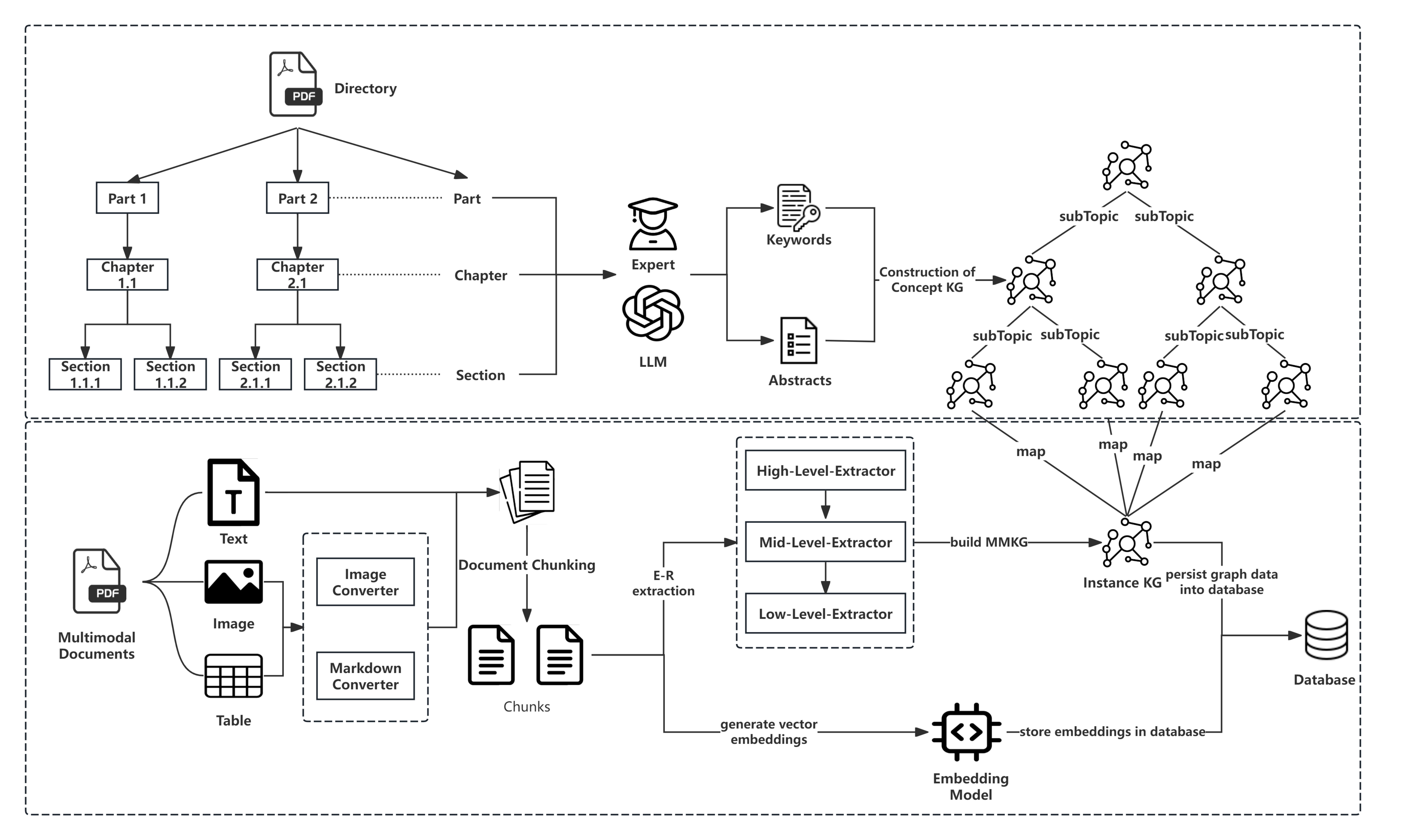} 
    \caption{Flow Diagram of DSKG Construction from Original Documents to MMKG} 
    \label{fig:third-image} 
\end{figure}

    As shown in Fig.\ref{fig:third-image}, the proposed DSKG construction process encompasses three stages: data preprocessing, construction of Concept KG, and construction of Instance KG, forming a full pipeline from raw domain-specific documents to multimodal knowledge graph.
    
	\subsubsection{Data Preprocessing.} This work employs the Mineru document mining tool and a hybrid OCR engine to perform pixel-level parsing of original PDFs, converting them into a standardized JSON-LD format for structured representation. Text is denoised for consistency, while tables are processed with algorithms to detect row and column structures and converted to Markdown format. Deep learning-based vision models extract key features from images and generate natural language descriptions for further processing. All outputs are unified into a standardized text format for knowledge extraction and graph construction. To address model input limitations, a paragraph-based text segmentation strategy is used, preserving semantic continuity. Tables and images are extracted as independent knowledge chunks for integration into the multimodal knowledge graph.

    \subsubsection{Concept KG Construction.} In domain-specific technical documents, the table of contents serves as both an organizational tool and a reflection of implicit knowledge and engineering semantics. This study proposes a method for constructing a concept knowledge graph driven by the document's structural hierarchy, leveraging existing semantic information for clear, hierarchical modeling. Elements such as Part, Chapter, and Section titles are treated as concept nodes, with each section's summary representing its semantic core. Supported by expert annotations and large language models, a network of concept subgraphs is built, linking overarching frameworks to detailed content and forming an initial concept knowledge graph with coherent logic.

	Based on chapter-level semantic abstracts, entity and relation extraction models are used to identify key concept associations. Hierarchical edge relations (e.g., subTopic, hasKeyword) are then defined through structured modeling, forming a Directed Acyclic Graph (DAG) of multi-level subgraphs. This concept KG preserves the original document's organizational logic and provides clear pruning paths for the downstream QA, supporting layer-by-layer retrieval based on the chapter structure. The final concept knowledge graph anchors both structural and semantic dimensions, enabling efficient alignment and unified representation for subsequent multimodal instance graph construction.

    \subsubsection{Instance KG Construction.} Building upon the modeling of the concept knowledge graph, this work addresses the heterogeneity and granularity differences of instance-level knowledge in domain-specific multimodal documents. To more effectively support the extraction and fusion of multimodal entities and relations, a structured chunking strategy is employed to segment documents into minimal semantic units, each annotated with corresponding chapter structures and contextual information. After this processing, each figure or table is treated as an independent semantic unit (chunk) and incorporated alongside its associated paragraph into the entity-relation extraction pipeline. This approach maintains modality diversity while enhancing structural clarity and ensuring consistency in knowledge graph modeling.

\begin{figure}
    \centering
    \includegraphics[width=0.5\linewidth]{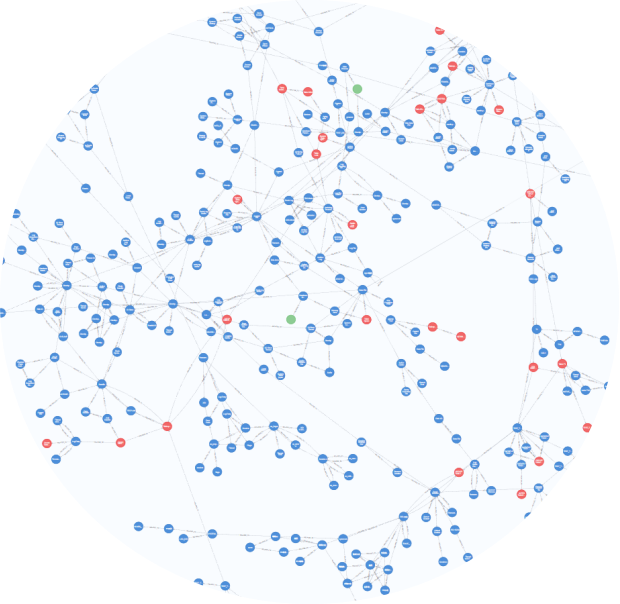} 
    \caption{Visualization of the DSKG} 
    \label{fig:fourth-image} 
\end{figure}

    This work designs a three-layered extraction agent architecture: High-Level, Mid-Level, and Low-Level Extractors. The High-Level Extractor identifies macro-level semantic relationships based on the document's hierarchy. The Mid-Level Extractor uses domain ontologies and context to extract core entities and operations. The Low-Level Extractor focuses on technical entities, particularly procedural, identificational, and parametric entities, while an attribute completion mechanism enriches entity properties. All extracted information is integrated with the concept KG subgraphs to construct a multimodal, hierarchical knowledge graph, as shown in Fig.\ref{fig:fourth-image}.

	\subsection{DSKG-Enhanced Retrieval}
\begin{figure}
    \centering
    \includegraphics[width=1\linewidth]{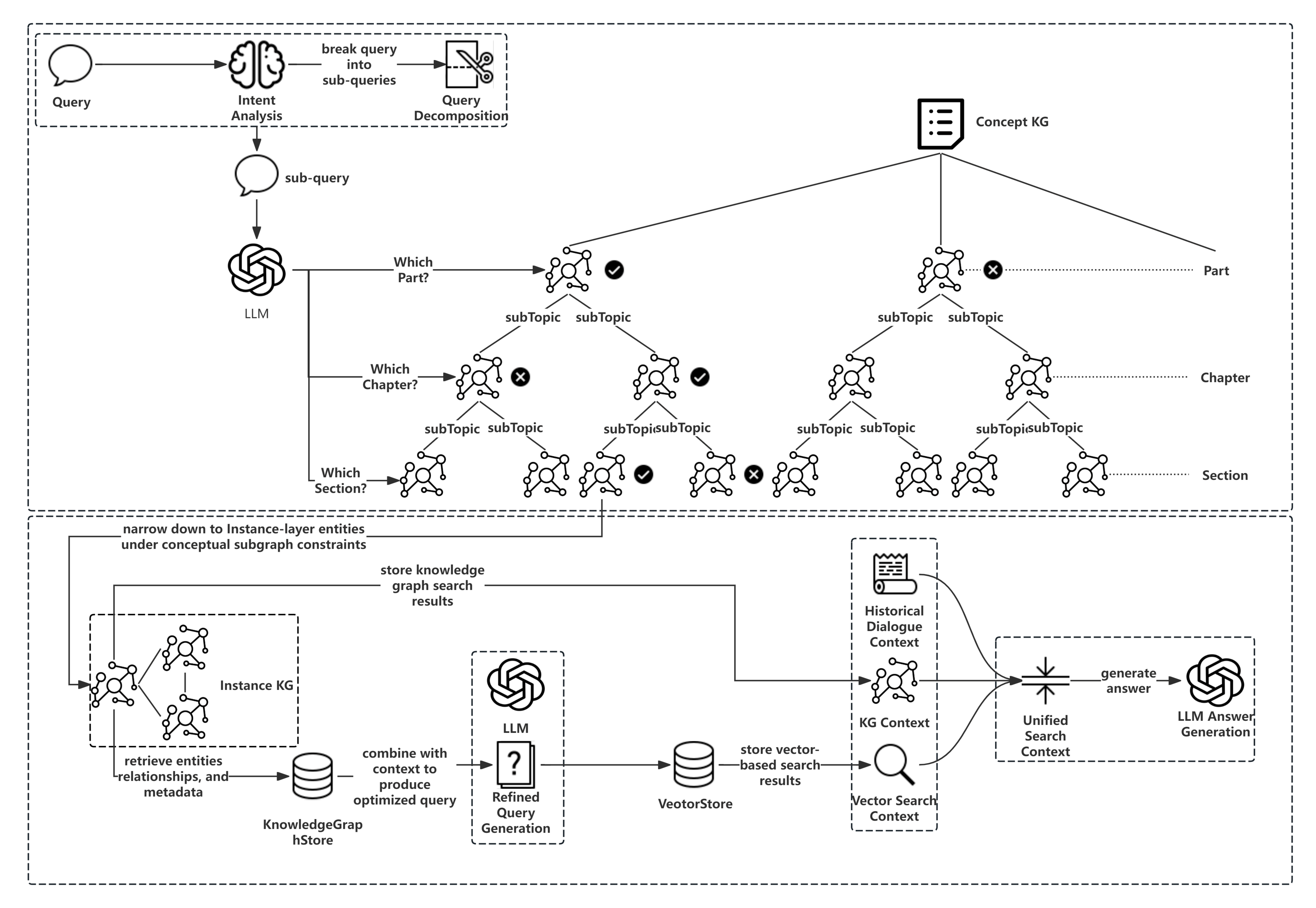} 
    \caption{DSKG-enhanced Retrieval Process} 
    \label{fig:fifth-image} 
\end{figure}

    As shown in Fig.\ref{fig:fifth-image}, the proposed DSKG-enhanced retrieval combines structured graph semantic modeling with unstructured corpus vector search, forming a multi-source retrieval-augmented generation framework that builds a knowledge acquisition path from structure to semantics and from concepts to instances.

    \subsubsection{Graph-guided Focusing.} In the graph-guided focusing stage, structured knowledge graphs are leveraged to precisely model query intent and spatially constrain the contextual semantics. The process first performs semantic decomposition of the original query and generates sub-queries, each corresponding to a different semantic sub-dimension.
    
    Based on this semantic modeling, a layer-wise pruning strategy is applied for hierarchical matching and conceptual constraint: at the concept KG, irrelevant parts are eliminated through semantic matching, and relevant chapters and sections are located, progressively narrowing the search space to enhance retrieval efficiency and accuracy; at the instance KG, the focused concept subgraph serves as a boundary to match structured entities, attributes, and relations, extracting instance-level knowledge closely related to the query target.After all sub-queries are completed, the retrieved results are merged into a knowledge graph subset, forming a structured semantic context highly aligned with the original query.
    
    \subsubsection{Vector Retrieval.} Upon completing the semantic focusing process guided by the knowledge graph, a vector-based retrieval mechanism is subsequently introduced to address the limitations of structured representations in terms of coverage and fine-grained detail. Specifically, based on the chapter nodes identified through graph pruning, corresponding document chunks are selected as the candidate set for vector retrieval, facilitating the alignment of semantics between the structured knowledge graph and the document embedding space.

    Two distinct prompt templates are dynamically generated from the structured semantic context of the graph.  Integrating the initial query with the graph's contextual semantics, we create a refined query and encode it into a high-dimensional vector for similarity search in the candidate set.Ultimately, the structured graph semantics, retrieved text fragments, and historical dialogue are then consolidated into a Unified Search Context, which serves as input for the language model to generate accurate, contextually relevant answers. This mechanism aligns structured knowledge with unstructured information, enhancing the precision and comprehensiveness of domain-specific question answering.

 \section{Experiment and Results}\label{Work4}
	\subsection{Experiment Setup}
	\subsubsection{Dataset.} The experimental data in this study come from domain-specific multimodal technical documentation in the database field, covering various aspects of database systems. With over 4,500 pages, the document includes text, tables, images, and code snippets, forming a multidimensional knowledge source. It is divided into seven main parts, containing 54 specialized chapters and approximately 900 knowledge units, creating a clear hierarchical structure for building a domain-specific knowledge graph. To evaluate the DSRAG framework's effectiveness in database question-answering tasks, domain experts designed 100 high-quality questions, spanning areas like database operations, configuration management, and performance optimization, with corresponding ground truths.

 \subsubsection{Environments.} The experimental environment setup is shown in Table~\ref{tab:experimental_config}, covering hardware configurations, software versions, and details of the various models used in the experiments.

\begin{table}[H]
\centering
\small 
\caption{Experimental Configuration Details}\label{tab:experimental_config}
\resizebox{\textwidth}{!}{
\begin{tabular}{|l|l|l|}
\hline
Category & Configuration & Details \\ 
\hline
\multirow{3}{*}{Hardware} 
& CPU & Intel Xeon Gold 6226R (32 cores) @ 2.90GHz \\
& Memory & 256 GB \\
& Graphics Card & NVIDIA RTX A5000 \\
\hline
\multirow{3}{*}{Software} 
& System & Ubuntu 22.04.5 LTS \\
& Software & CUDA 12.8 \\
&  & Python 3.11 \\
\hline
\multirow{5}{*}{Models} 
& Generation Model & GPT-4o-mini \\
& Graph Construction Model & GPT-4o-mini \\
& Evaluation Model & GPT-4o \\
& Embedding Model & text-embedding-3-small \\
& Reranking Model & jina-reranker-v2-base \\
\hline
\end{tabular}
}
\end{table}

\subsubsection{Baselines.} 
To validate the key role of domain knowledge enhancement in complex question answering tasks and explore its effectiveness in improving answer factuality, contextual relevance, and cross-modal reasoning ability, this study compares the following three baseline methods based on technical diversity and domain adaptability:

\begin{itemize}
    \item NativeRAG. A baseline RAG framework that combines general retrieval and generation modules without incorporating knowledge graph.
    \item TiDB AutoFlow. A graph-enhanced RAG framework proposed by PingCAP, aiming to integrate structured domain knowledge into the RAG process to enhance factual consistency and semantic alignment.
    \item RAGFlow. A multimodal RAG engine developed by InfiniFlow, supporting heterogeneous data sources and employing hybrid retrieval strategies to enhance context coverage.
\end{itemize}

\subsubsection{Metrics.} We evaluated the performance using three metrics: Faithfulness, Answer Relevancy, and Contextual Precision. Faithfulness is calculated by determining the proportion of context-inferred statements in the generated answer compared to the total number of statements, assessing the factual consistency of the answer with the context. Answer Relevancy typically uses cosine similarity between the answer and the query to measure how well the answer matches the query.Contextual Precision assesses whether factually relevant entries are ranked higher within the retrieved context.These metrics together ensure a comprehensive evaluation of answer quality, factual grounding, and retrieval effectiveness.

	\subsection{ Performance Comparison}
    \subsubsection{Overall Performance.}As shown in Table~\ref{tab:performance_comparison}, we compared the proposed method with other baseline methods. The experimental results demonstrate that the DSRAG method significantly outperforms all baseline methods in terms of all evaluation metrics, including Faithfulness, Answer Relevancy, and Contextual Precision. Specifically, the method excels in generating factually consistent answers, providing highly relevant user responses, and accurately identifying and ranking context from the knowledge base. This ensures the precision and relevance of the answers, validating its effectiveness in domain-specific question answering tasks.It also highlights the significance of combining domain-specific multimodal knowledge graphs with retrieval-augmented generation techniques to improve the performance of intelligent question-answering tasks.

\begin{table}[H]
    \centering
    \small 
    \caption{Performance Comparison between DSRAG and Baseline Methods}
    \label{tab:performance_comparison}
    \setlength{\tabcolsep}{5pt} 
    \renewcommand{\arraystretch}{1.2}
    \begin{tabular}{|c|c|c|c|}
        \hline
        Method  
        & Faithfulness  
        & Answer Relevancy  
        & Context Precision \\
        \hline
        Naive RAG        & 0.63 & 0.58 & 0.55 \\
        \hline
        TiDB AutoFlow    & 0.72 & 0.65 & 0.62 \\
        \hline
        RAGFlow          & 0.69 & 0.67 & 0.70 \\
        \hline
        \textbf{DSRAG}   & \textbf{0.83} & \textbf{0.87} & \textbf{0.85} \\
        \hline
    \end{tabular}
\end{table}

    \subsubsection{Ablation Study.} To systematically evaluate the contribution of different components within the proposed framework, we conducted an ablation study under three experimental settings: (1) the baseline retrieval-augmented generation (RAG) method without knowledge graph integration; (2) RAG enhanced with only the IKG; and (3) the complete DSRAG framework, incorporating both the CKG and IKG (DSKG). By incrementally integrating different knowledge graph components, we assessed their impact on the overall performance of domain-specific question answering tasks. As summarized in Table~\ref{tab:ablation_study},  the complete DSRAG framework achieved the best results across all metrics, with a faithfulness score of 0.83 and an answer relevance score of 0.87, significantly outperforming the other configurations. In contextual precision, DSRAG also reached 0.85, demonstrating the critical role of comprehensive structured knowledge in improving factual consistency, semantic relevance, and retrieval accuracy.
\begin{table}[H]
    \centering
    
    \caption{Ablation Study on RAG Configurations with Different Levels of Knowledge Graph Incorporation}
    \label{tab:ablation_study}
    \setlength{\tabcolsep}{5pt}
    \renewcommand{\arraystretch}{1.2}
    \begin{tabular}{|c|c|c|c|}
        \hline
        Method  
        & Faithfulness  
        & Answer Relevancy  
        & Context Precision \\
        \hline
        RAG             & 0.68 & 0.70 & 0.58 \\
        \hline
        IKG-RAG & 0.72 & 0.80 & 0.69 \\
        \hline
        \textbf{DSKG-RAG} & \textbf{0.83} & \textbf{0.87} & \textbf{0.85} \\
        \hline
    \end{tabular}
\end{table}

\section{Conclusion}\label{Work5}
	This paper introduces DSRAG, a multimodal knowledge graph-driven retrieval-augmented generation (RAG) framework tailored for domain-specific tasks, enhancing response accuracy and domain adaptability. DSRAG constructs a multimodal knowledge graph integrating both concept and instance KGs, optimizing retrieval and generation through semantic pruning and structured subgraph retrieval. Experimental results demonstrate that DSRAG outperforms traditional baselines in key metrics like faithfulness, answer relevance, and contextual accuracy. Ablation studies confirm the complementary benefits of combining concept and instance KGs. This work offers a novel approach to intelligent question answering in domain-specific fields and highlights the potential of integrating multimodal knowledge graphs with RAG. Future work can explore advanced graph structures and cross-modal fusion for more complex tasks.

\bibliographystyle{splncs04}
\bibliography{references}
\end{document}